\documentclass{svproc}
\pdfoutput=1
\usepackage[bulletsep]{collref}

\renewcommand{\geq}{\geqslant}

\usepackage{url}
\usepackage{bbm}

\usepackage{amssymb,graphicx}

\begin{document}
\mainmatter              
\title{Integrable Holographic Defect CFTs}
\titlerunning{Integrable Holographic dCFTs}  
%
\author{Charlotte Kristjansen\inst{1} and Konstantin Zarembo\inst{1,}\inst{2}}

\authorrunning{C.\ Kristjansen \& K.\ Zarembo} 
%
\tocauthor{C.\ Kristjansen and K. \ Zarembo}

\institute{Niels Bohr International Academy, Niels Bohr Institute, University of Copenhagen\\
Blegdamsvej 17, DK-2100 Copenhagen \O, Denmark 
\email{kristjan@nbi.dk,zarembo@nbi.ku.dk }  
\vspace*{0.4cm}
\and
Nordita, KTH Royal Institute of Technology and Stockholm University, \\
Hannes Alfv\'ens v\"ag 12, SE-106 91 Stockholm, Sweden \\
\email{zarembo@kth.se}}


\maketitle              
\begin{abstract}
We review a class of integrable, supersymmetric defect conformal field theories which have holographic duals in the form of probe brane models. 
Our main examples are defect versions of  ${\cal N}=4$ SYM and  ABJM theory, both involving a domain wall 
with Nahm pole boundary
conditions, and the case of a 't Hooft line embedded in  ${\cal N}=4$ SYM.
 The  field theory defect respectively the probe D-brane can be described as 
an integrable boundary state of the integrable system underlying the AdS/CFT correspondence.

\end{abstract}
\section{Introduction}
Holography and integrability appear to be closely tied to each other with the most well established examples being the 
AdS$_{5}$/CFT$_4$ and the  AdS$_{4}$/CFT$_3$ correspondence where in both cases the planar spectral problem is 
described in terms of an integrable spin chain~\cite{Beisert:2010jr}.
 Supersymmetry and conformal symmetry  seem  to be key prerequisites for this connection. From the original AdS/CFT systems one can engineer set-ups which conserve a subset of the supersymmetries and likewise carry conformal symmetry only in a certain subspace. We will consider configurations where one introduces a flat defect in the field theory 
 and correspondingly a probe brane in the string theory in such a way that the entire system is 1/2 BPS and conformal symmetry
 is conserved on the defect, leading to a defect CFT on the field theory side. In particular, we will highlight three models where integrability is preserved after the reduction in conformal- and super-symmetry as a consequence of 
the defect/probe brane being identified with an integrable boundary state of the underlying spin chain. Two of these models
have defects of co-dimension one and are built from respectively ${\cal N}=4$ SYM theory and ABJM theory by the introduction of a domain wall at which Nahm pole boundary conditions are imposed on a part of the field components~\cite{Gaiotto:2008sa,deLeeuw:2015hxa,Terashima:2008sy,Kristjansen:2021abc}. The third model involves 
a defect of co-dimension three and is defined by a 't Hooft line embedded in  ${\cal N}=4$ SYM theory~\cite{Kapustin:2005py,Kristjansen:2023ysz}. The three models are treated in respectively section~\ref{SYM}, \ref{ABJM}
and~\ref{tHooft}, and we finish with a summary and outlook in section~\ref{conclusion}. In addition, we
have included an appendix which briefly reviews the concepts needed from the language of spin chain integrability.

\section{The Nahm pole defect in ${\cal N}=4$ SYM \label{SYM}}

\begin{figure}[t]
 \includegraphics[width=8.5cm,angle=270,origin=c] {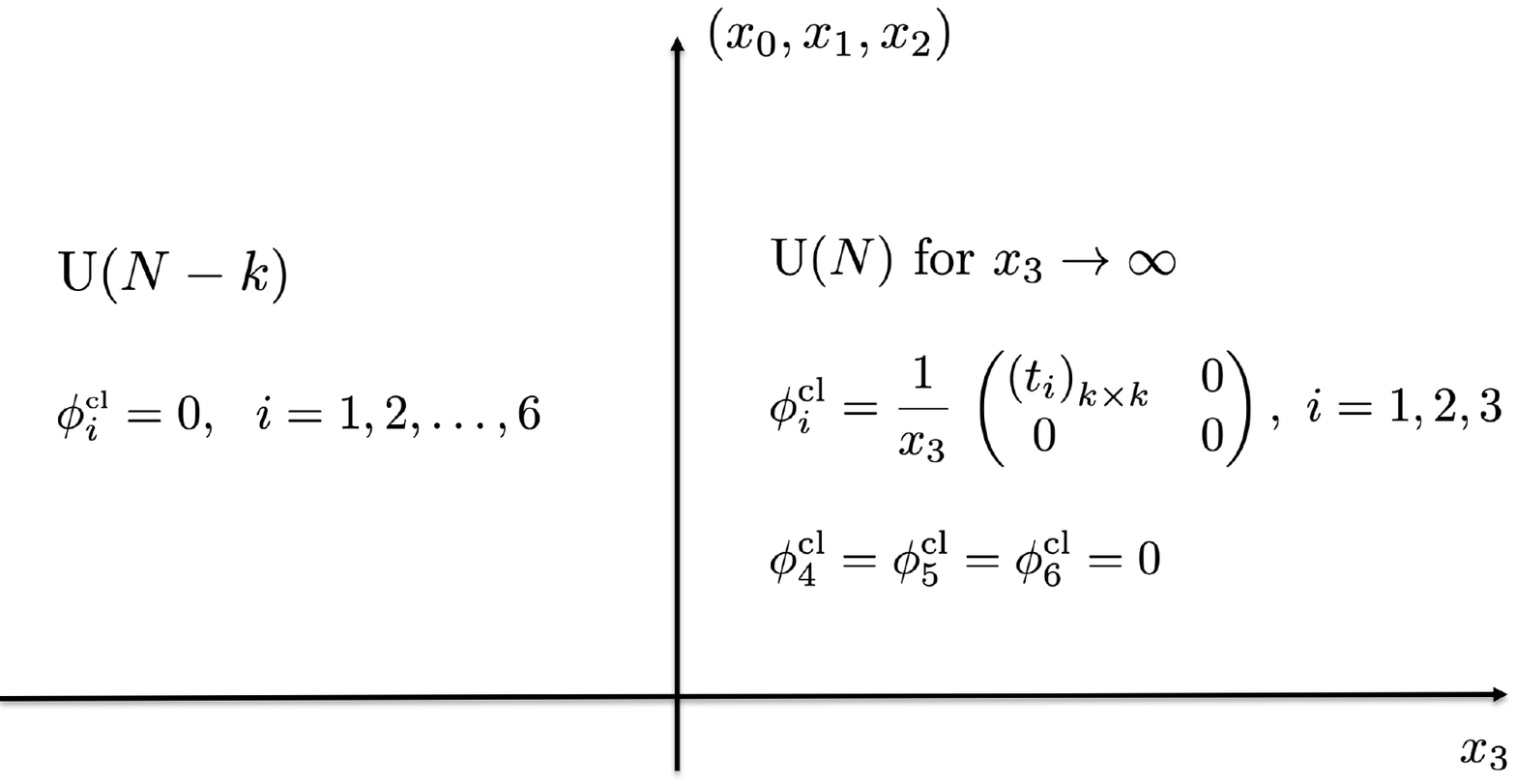}
\vspace*{-3.2cm}
 \caption{\label{NahmSYM}The Nahm pole defect in ${\cal N}=4$ SYM}
\end{figure}

Supersymmetric boundary conditions in ${\cal N}=4$ SYM have been extensively studied,   e.g. in~\cite{Gaiotto:2008sa,Gaiotto:2008sd}, and include  Nahm pole boundary conditions~\cite{Nahm:1979yw,Diaconescu:1996rk,Constable:1999ac}  where three out of the six
scalar fields of the theory diverge in a way compatible with conformal symmetry as the boundary is approached.
With these boundary conditions as the starting point one can construct a defect CFT version of ${\cal N}=4$ SYM theory which
has an explicit holographic dual. More precisely one glues together two copies of ${\cal N}=4$ SYM with different ranks of the gauge group, $N$ and $N-k$, along a co-dimension one surface. The difference in the rank of the gauge group on the 
two sides of the defect is accounted for by Nahm pole boundary conditions on certain components of the
classical fields, i.e.
\begin{equation}
\label{eq:vevs}
\phi_i^\mathrm{cl} = \frac{1}{x_3}t_i \oplus 0_{(N-k)\times(N-k)}\,,\hspace{0.5cm}  x_3>0, \hspace{0.5cm} i=1,2,3,
\end{equation}
where the $t_i$'s generate a $k$-dimensional irreducible representation of SU$(2)$, i.e.
\begin{equation}
[t_i,t_j]=\epsilon_{ijk} t_k. \label{commutation}
\end{equation}
\`{A} priori, we do not impose any  boundary conditions on the fields $\phi_{4,5,6}$ at the defect
In the holographically dual string theory picture  the defect corresponds to a single probe $D5$-brane on which $k$ of the $N$
D3-branes of the standard AdS/CFT construction end. The configuration is summarized in Fig.~\ref{NahmSYM}. 
It constitutes a 1/2 BPS set-up preserving the residual OSp$(4|4)$ symmetry. For a more detailed description we refer to~\cite{deLeeuw:2017cop,deLeeuw:2019usb,Linardopoulos:2020jck}.

\begin{figure}[t]
\begin{center}
 \includegraphics[width=3.2cm] {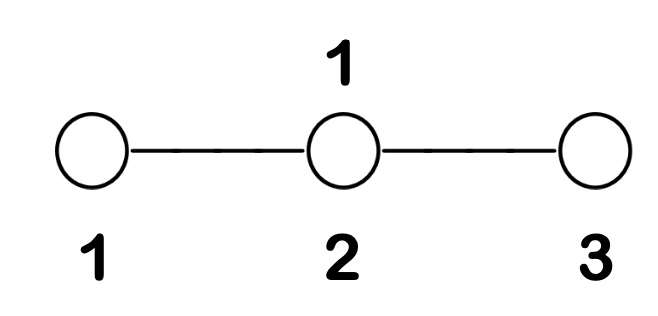}
\end{center}
 \caption{\label{SO6Dynkin} The Dynkin diagram of $SO(6)$. Above the nodes are the weights of the relevant spin representation while the numbers below the nodes are the labels of the $Q$-functions.}
\end{figure}

Being a defect CFT the Nahm pole version of ${\cal N}=4$ SYM theory is endowed with additional non-trivial conformal data as compared to a standard CFT. These include one-point functions, two-point functions of operators with unequal conformal
dimensions, and correlation functions involving bulk and boundary fields. 
An interesting analysis of the operator algebra can be found in~\cite{Dedushenko:2020vgd,Dedushenko:2020yzd}.
In the present discussion we will concentrate on
one-point functions which are bound to take the form
\begin{equation}
\langle {\cal O}_{\Delta}\rangle=\frac{C}{|x_3|^\Delta},
\end{equation}
with $\Delta$ being the conformal dimension.
The single trace conformal operators built entirely from scalars are at one-loop order in one-to-one correspondence with the eigenstates, 
$|\bf{u}\rangle$, of an integrable SO$(6)$ spin chain~\cite{Minahan:2002ve}
\begin{equation}\label{so6-operators}
 \mathcal{O}=\Psi ^{I_1\ldots I_L}\mathop{\mathrm{tr}}\phi _{I_1}\ldots \phi _{I_L} \sim |\bf{u}\rangle
\end{equation}
The one-point function of such an operator is at leading order obtained simply by replacing each quantum field with its classical value. This process can be implemented
by computing the overlap of the Bethe eigenstate with a matrix product state defined as follows
\begin{equation}\label{MPSk}
\mbox{MPS}_k^{\,\, I_1 I_2\ldots I_L}=\mathrm{tr}\, t_{I_1}t_{I_2}\ldots t_{I_L},
\end{equation}
with the understanding that $t_4=t_5=t_6=0$.
More precisely, up to a field theoretical pre-factor the overlap $C$ can be expressed as~\cite{deLeeuw:2015hxa,Buhl-Mortensen:2015gfd}
\begin{equation}
C_k\left(\bf{u}\right)= \frac{\langle \mathrm{MPS}_k \left|\bf{u}\rangle\right.} 
{\left\langle  \bf{u}\left|\bf{u}\right.\right\rangle^{\frac{1}{2}}}, \hspace{0.5cm} k\geq 2.
\end{equation} 
In the special case $k=1$ one can still construct a defect CFT in such a way that the configuration becomes 1/2 BPS by appropriately imposing Dirichlet respectively Neuman
boundary conditions on the field components outside the $(N-k)\times (N-k)$ block for $x_3\rightarrow 0_+$~\cite{deLeeuw:2017dkd}. In this case the matrix product state must be replaced by a valence bond state (i.e.\ a two-site product state), more precisely~\cite{Kristjansen:2020mhn}
\begin{equation}
\mbox{VBS}^{\,\, I_1 I_2\ldots I_L}= K^{I_1I_2}\ldots K^{I_{L-1}I_L},
 \hspace{0.5cm} K^{IJ} =2 \sum_{I,J=1}^3 \delta^{IJ}-2\sum_{I,J=4}^6\delta^{IJ}.
\end{equation}
Using the language and tools of integrability the overlap between the Bethe eigenstate of the integrable $SO(6)$ spin chain and this valence bond state was determined to be~\cite{DeLeeuw:2019ohp}
\begin{equation}
 \frac{\left\langle {\rm VBS}\right.\!\left|\mathbf{u} \right\rangle}{\left\langle \mathbf{u}\right.\!\left|\mathbf{u} \right\rangle^{\frac{1}{2}}}
 =2^{-L}\,\sqrt{\frac{\mathop{Q_1(0)Q_2(0)Q_3(0)}}{Q_1\left(\frac{i}{2}\right)Q_2\left(\frac{i}{2}\right)Q_3\left(\frac{i}{2}\right)}\mathrm{Sdet}G}
 \,.
\end{equation}
Here and in the later overlap formulas $Q_a(x)$ is the Baxter polynomial involving the Bethe roots associated with the $a$'th node of the Dynkin diagram chosen to
represent the underlying symmetry algebra,
and $\mbox{SdetG}$ is the superdeterminant of the so-called Gaudin matrix which encodes the norm of the Bethe eigenstate. We notice, in particular, that the factor 
SdetG contains no reference to the defect.
In Fig. \ref{SO6Dynkin} we show the $SO(6)$ Dynkin diagram and our convention for the numbering of the nodes.  
Furthermore, we have included an appendix with a brief introduction to the language of integrability.

\begin{figure}[h]
\vspace*{-3.0cm}
 \includegraphics[width=8.5cm,angle=270,origin=c] {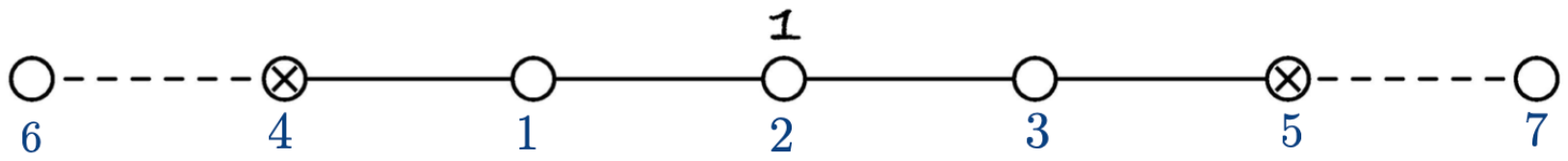}
 \vspace*{-5.0cm}
 \caption{\label{SYMDynkin}One of the Dynkin diagrams of $PSU(2,2|4)$ and its associated overlap formula in eqn.~(\ref{VBSgeneral}).}
\end{figure}

The overlap with the matrix product state can likewise be found in a closed form~\cite{DeLeeuw:2018cal}.
\begin{equation}
C^{SO(6)}_k =
\sqrt{
\frac{Q_2(0)Q_2(\frac{i}{2})Q_2(\frac{ik}{2})Q_2(\frac{ik}{2})}{{Q}_3(0){Q}_3(\frac{i}{2}){Q}_1(0){Q}_1(\frac{i}{2})}
} \cdot \mathbb{T}_{k-1}(0) \cdot \sqrt{\mbox{Sdet} G},\label{SO6formula}
\end{equation}
where
\begin{equation}\label{TSO(6)}
\mathbb{T}_n(x) = \sum_{a=-\frac{n}{2}}^{\frac{n}{2}}(x+ia)^L \frac{Q_3(x+ia)Q_1(x+ia)}{Q_2(x+i(a+\frac{1}{2})) Q_2(x+i(a-\frac{1}{2}))}.
\end{equation}
Starting from the overlap with the valence bond state in the SO$(6)$ case and requiring that the overlap is covariant under fermionic dualities one can deduce the dependence
of the leading order overlap on Baxter $Q$-functions for the full defect ${\cal N}=4$ SYM theory for any of the possible Dynkin diagrams of its underlying symmetry 
algebra $\mathfrak{psu}(2,2|4)$~\cite{Kristjansen:2020vbe}.
Here we give the formula for the overlap in the grading depicted in Fig.~\ref{SYMDynkin}:
\begin{equation} \label{VBSgeneral}
\frac{\left\langle {\rm VBS}\right.\!\left|\mathbf{u} \right\rangle}{\left\langle \mathbf{u}\right.\!\left|\mathbf{u} \right\rangle^{\frac{1}{2}}}= 2^{-L}
\sqrt{\frac{Q_1(0)Q_3(0)Q_4(0)Q_5(0)Q_7(0)}{Q_2(0)Q_2(\frac{i}{2})Q_4(\frac{i}{2})Q_6(0)Q_6(\frac{i}{2})} \,S \det G}
\end{equation}
In the so-called alternating grading, it is possible to derive an expression for the the overlap with the valence bond state
valid to any loop order in the asymptotic limit~\cite{Gombor:2020kgu,Komatsu:2020sup,Gombor:2020auk} (i.e.\ excluding wrapping corrections~\cite{Ambjorn:2005wa}).  The derivation is based on applying bootstrap arguments to the boundary 
reflection matrix of either the spin chain~\cite{Gombor:2020kgu,Gombor:2020auk} or the string sigma model involved~\cite{Komatsu:2020sup}. It would be interesting to extend the analysis to including wrapping corrections e.g.\ by
 the thermodynamical Bethe ansatz approach. From the string theory perspective, it can be argued that the model is
 integrable in the classical limit~\cite{Linardopoulos:2021rfq}.

\section{The Nahm pole defect in ABJM theory\label{ABJM}}

\begin{figure}[t]
 \includegraphics[width=8.4cm,angle=270,origin=c] {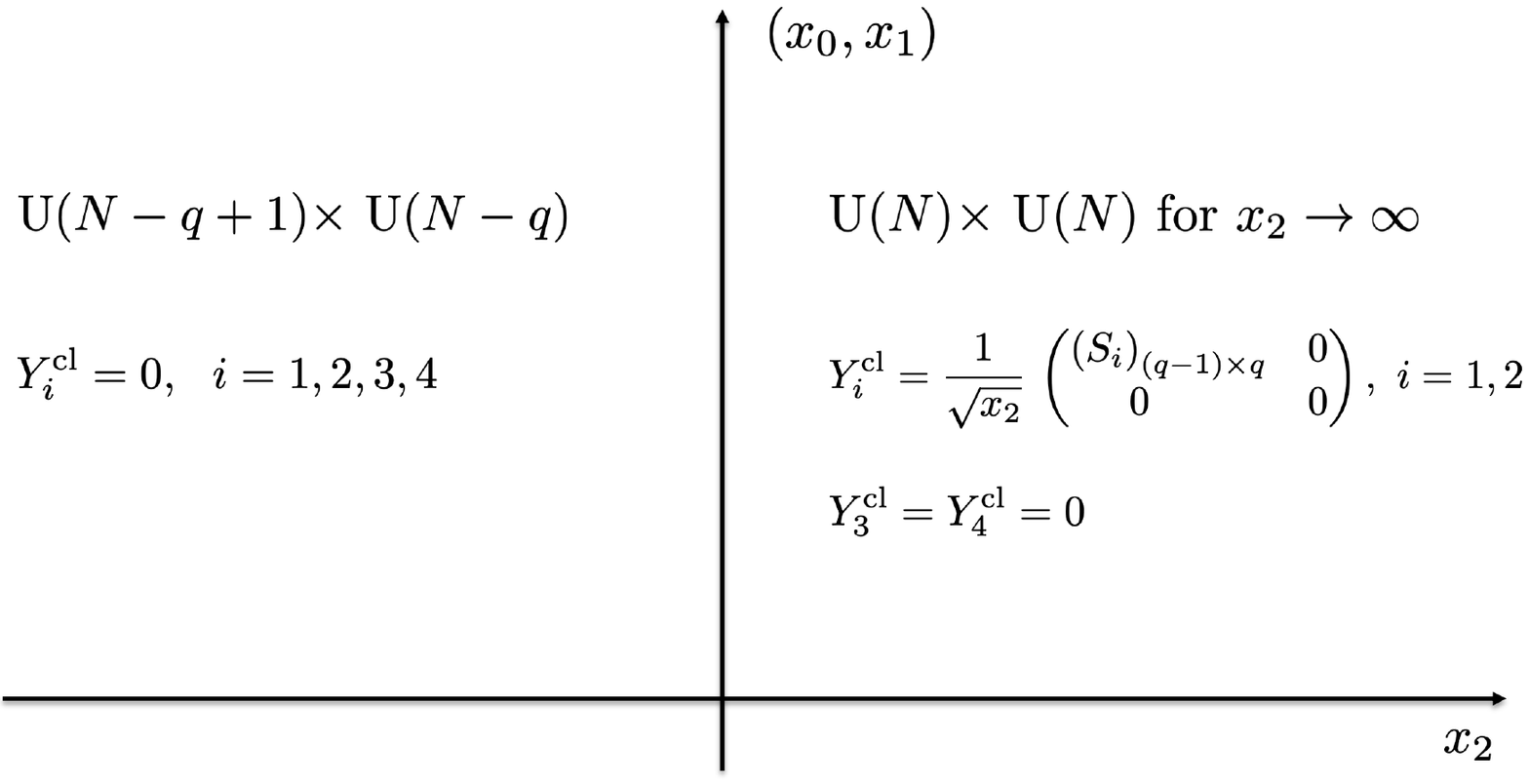}
 \vspace*{-3.3cm}
 \caption{\label{NahmABJM}The Nahm pole defect in ABJM theory}
\end{figure}

 The understanding of supersymmetric boundary conditions is much less developed for ABJM theory than for ${\cal N}=4$ SYM. Possible Nahm pole
boundary conditions for the four complex scalar fields of ABJM theory were identified in~\cite{Terashima:2008sy} and can be used to define
a domain wall version of ABJM theory in analogy with the construction for ${\cal N}=4$ SYM. The resulting defect set-up is illustrated in figure~\ref{NahmABJM}. 
It conserves half of the supersymmetries with the remaining symmetry group being $OSp(3|2)\times OSp(3|2)\times U(1)$. Two out of the four complex fields have a non-trivial Nahm pole configuration
for $x_2>0$ where we notice that the $x_2$ scaling is now exactly as needed to produce a dCFT in three dimensions and where the  two rectangular matrices $S_i$ are given by
\begin{equation}\label{Terashima2}
 S^1_{ij}=\delta _{i,j-1}\sqrt{i},\qquad S^2_{ij}=\delta _{ij}\sqrt{q-i}\,,\qquad i=1,\ldots ,q-1\qquad j=1,\ldots ,q.
\end{equation}
In the dual string theory description the defect corresponds to a single probe D4-brane on which $q$ of the $N$ D2-branes
of the ABJM construction end. The matrices $S^i$ are related to $SU(2)$ representation matrices in the following way:
Defining the composite classical field
\begin{equation}
 \Phi ^\alpha _{\hphantom{\alpha }\beta }=Y^\alpha Y^\dagger _\beta\equiv \Phi ^i\sigma ^\alpha _{i\,\beta }
  + \Phi \delta ^\alpha _{\hphantom{\alpha }\beta },
\end{equation}
one can show that~\cite{Kristjansen:2021abc} 
\begin{equation}
\Phi ^i=\frac{t^i}{x_2}\,, \hspace{0.5cm} \Phi =\frac{q\mathbbm{1}}{2x_2}.
\end{equation}
where the $t_i$ fulfill the $SU(2)$ commutation relations~(\ref{commutation}). 

\begin{figure}
\begin{center}
 \includegraphics[width=3.2cm] {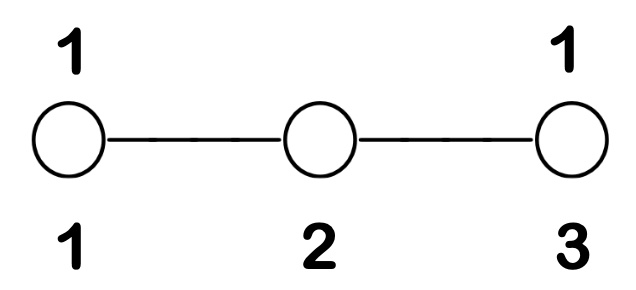}
 \end{center}
\caption{\label{SU4Dynkin}The Dynkin diagram of $SU(4)$ with the associated overlap formula given in eqn.~(\ref{overlap}). Above the nodes are the weights of the relevant spin representation while the numbers below the nodes are the labels of the $Q$-functions.}
\end{figure}

The single trace conformal operators of ABJM theory built entirely from scalars take the following form and are in one-to-one correspondence with the eigenstates, $|\bf{u}\rangle$, of an alternating $SU(4)$ spin chain~\cite{Minahan:2008hf}
\begin{equation}
 \mathcal{O}=\Psi _{A_1\,\ldots\, A_{2L-1}}^{\hphantom{A}A_2\,\ldots\, A_{2L}}\mathop{\mathrm{tr}}Y^{A_1}Y^\dagger _{A_2}\ldots Y^{A_{2L-1}}Y^\dagger _{A_{2L}}.
\end{equation}
To compute the one-point functions of these operators in the Nahm pole background (for $x_2>0$) one needs to evaluate
the overlap of the wavefunction $\Psi$ with the wave function describing an appropriate matrix product state~\cite{Kristjansen:2021abc}, i.e.
\begin{equation}
{\rm MPS_q} _{\hphantom{A}A_2\,\ldots\, A_{2L}}^{A_1\,\ldots\, A_{2L-1}}=\mathop{\mathrm{tr}}S^{A_1}S^\dagger _{A_2}\ldots S^{A_{2L-1}}S^\dagger _{A_{2L}}.
\end{equation}
We notice that for the special case $q=2$ the matrix product state degenerates and becomes a valence bond state
\begin{equation}
 {\rm VBS}^{\hphantom{\alpha}\alpha _1\,\ldots\, \alpha _{2L-1}}_{\,\,\hphantom{\alpha }\alpha _2\,\ldots\, \alpha _{2L}}={\rm MPS}^{\hphantom{2\,\,}\alpha _1\,\ldots\, \alpha _{2L-1}}_{2\,\,\hphantom{\alpha }\alpha _2\,\ldots\, \alpha _{2L}}=
 \delta _{\hphantom{\alpha }\alpha _2}  ^{\alpha_1  }\,\ldots \,
 \delta ^{\alpha _{2L-1}}  _{\hphantom{\alpha }\alpha_{2L}  }.
\end{equation}
The overlap between the valence bond state and the eigenstates of the alternating $SU(4)$ spin chain was 
found in~\cite{Gombor:2021hmj}: 
\begin{equation}\label{overlap}
 \frac{\left\langle {\rm VBS}\right.\!\left|\mathbf{u} \right\rangle}{\left\langle \mathbf{u}\right.\!\left|\mathbf{u} \right\rangle^{\frac{1}{2}}}
 =2^{-L}\,{Q_2(i)}\sqrt{\frac{\mathop{\mathrm{Sdet}}G}{Q_2(0)Q_2\left(\frac{i}{2}\right)}}
 \,.
\end{equation}
where the relevant Dynkin diagram and associated spin representation labels can be seen in Fig.\ \ref{SU4Dynkin}.
By an iterative argument including various assumptions on factorization this overlap formula can be generalized to higher values of $q$~\cite{Gombor:2022aqj}:
\begin{eqnarray}
\frac{\langle\mathrm{MPS}_{q}|\mathbf{u}\rangle}{\sqrt{\langle\mathbf{u}|\mathbf{u}\rangle}}&=&
\sum_{k=1} ^{q-1}
\frac{Q_{1}(\frac{i}{2})Q_{1}(i(q-\frac{1}{2}))}{Q_{1}(i(k-\frac{1}{2}))Q_{1}(i(k+\frac{1}{2}))}\,\, \, \times \nonumber \\
&&\left(\frac{k}{2}\right)^L\!\!\!Q_{2}(ik)\sqrt{\frac{\mathop{\mathrm{Sdet}}G}{{Q}_{2}(0){Q}_{2}(i/2)}}.\label{eq:general_form}
\end{eqnarray}
Invoking again invariance of the formula~(\ref{overlap}) under fermionic duality transformations one can deduce the form of 
the overlap with the valence bond state for any of the five possible Dynkin diagrams for the algebra $Osp(6|4)$~\cite{Kristjansen:2021abc}. 
Here we give the formula for the diagram
which has the $SU(4)$ Dynkin diagram above as a sub-diagram (Fig.~\ref{DynkinABJM}): 
\begin{figure}[h]
\hspace*{4.2cm} \includegraphics[height=3.6cm] {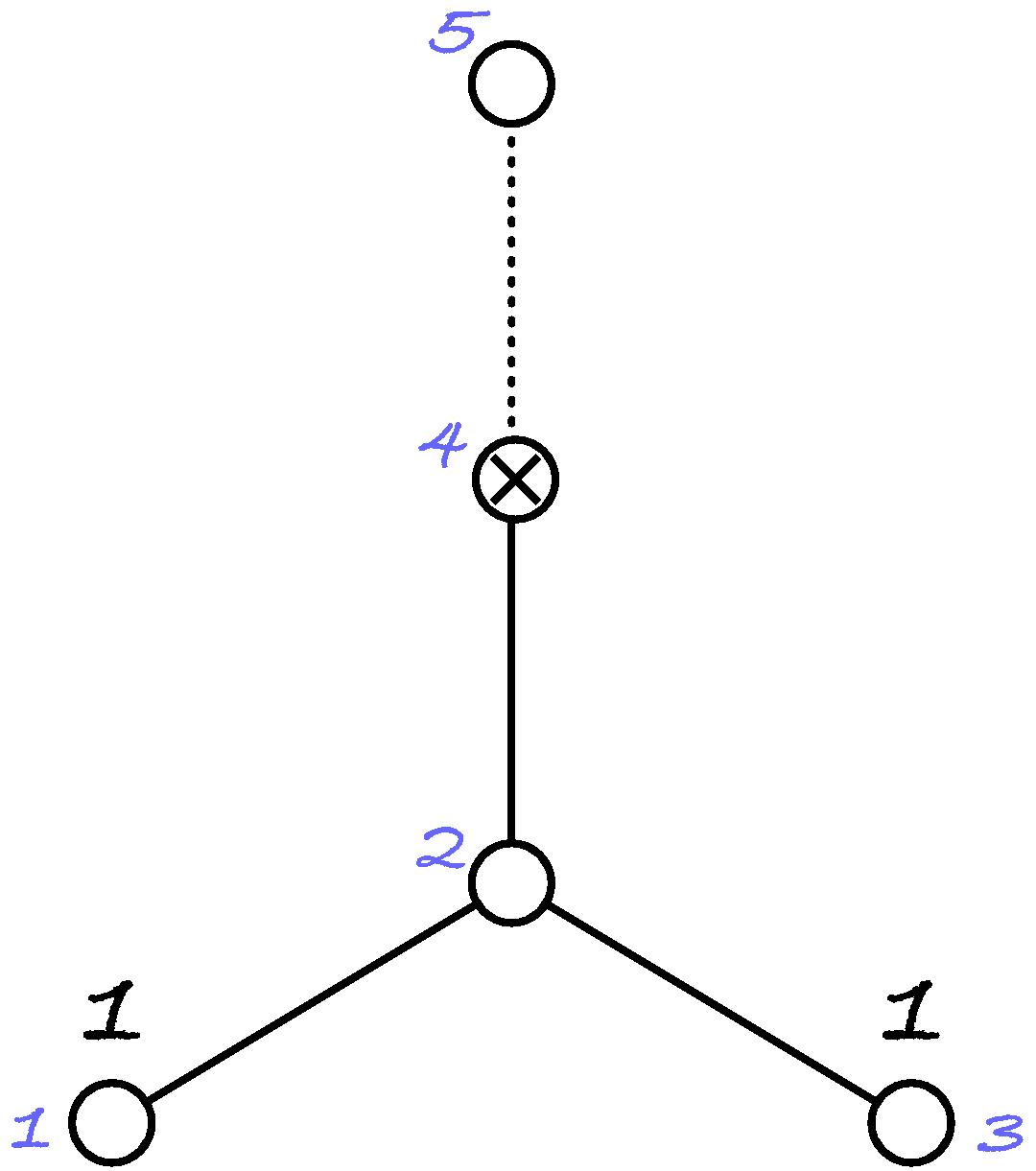}
 \caption{\label{DynkinABJM} One of the possible Dynkin diagrams of $\mbox{Osp}(6|4)$ 
with the associated overlap formula given in eqn.~(\ref{ABJMoverlap}).}
\end{figure}
\begin{equation}\label{ABJMoverlap}
 \frac{\left\langle {\rm VBS}\right.\!\left|\mathbf{u} \right\rangle}{\left\langle \mathbf{u}\right.\!\left|\mathbf{u} \right\rangle^{\frac{1}{2}}}
 = 2^{-L}\,{Q_2(i)}\sqrt{\frac{Q_4(0) Q_5(0)}{Q_2(0)\,Q_2\left(\frac{i}{2}\right)Q_5\left(\frac{i}{2}\right)}
 \mathop{\mathrm{Sdet}}G} 
 \,.
\end{equation}
Unlike what is the case for ${\cal N}=4$ SYM the overlap formula has not yet been extended to higher loop orders. The
dual string theory can be argued to be integrable in the classical limit~\cite{Linardopoulos:2022wol}.

\section{'t Hooft Loops\label{tHooft}}

The 't~Hooft loop is a line defect of co-dimension three.
Such defects describe point particles moving along predefined trajectories while interacting with the quantized gauge fields. Perhaps the best known example is the Wilson loop whose expectation value measures the response to an external color charge. Here we consider 't~Hooft loops \cite{Hooft:1977hy} which correspond to point-like magnetic monopoles. In the $\mathcal{N}=4$ SYM theory Wilson and 't~Hooft loops are interchanged by S-duality, an exact quantum counterpart to the electric-magnetic symmetry of electromagnetism \cite{Montonen:1977sn}.

The 't~Hooft loop is a disorder operator defined by singular boundary conditions on the field. For the straight line in the time direction,
\begin{eqnarray}\label{monopole}
 F_{ij}&\stackrel{r\rightarrow 0}{\simeq }&q\varepsilon _{ijk}\frac{x^k}{r^3},
\nonumber \\
\phi _I&\stackrel{r\rightarrow 0}{\simeq }&q\,\frac{n_I}{r}\,,
\end{eqnarray}
where $n_I$ is a unit six-vector defining the R-symmetry orientation of the monopole,
and $q=\mathop{\mathrm{diag}}(q_1,\ldots , q_N)$ is the diagonal embedding of the monopole charges in the $U(N)$ gauge group. For simplicity we consider the case of just one non-zero charge $q_1\equiv q$ and all other $q_i=0$. Only the $11$ element of the background field is then non-trivial (cf.~fig.~\ref{NahmSYM}). The $U(N)$ gauge symmetry gets broken to $U(N-1)\times U(1)$, with the off-diagonal components $\phi _{1i}$ and $\phi _{i1}=\phi _{1i}^*$ charged under the $U(1)$ subgroup where the monopole resides.

The path integral computed with these boundary conditions defines the expectation value and correlators of the 't~Hooft loop. Just like the Nahm solution for the domain wall, the monopole (\ref{monopole}) can be regarded as the classical background field.
In $\mathcal{N}=4$ SYM quantum corrections to the external field should vanish to all orders. The scalar coupling of the 't~Hooft loop is essential for that  \cite{Kapustin:2005py,Gomis:2009ir}:  with this choice the monopole preserves half of the supersymmetry. Together with the residual conformal symmetries of the line the unbroken supercharges form the $\mathfrak{osp}(4^*|4)$ superalgebra \cite{Liendo:2016ymz}. 

The AdS dual of the 't~Hooft line is a D1-brane in the bulk of $AdS_5\times S^5$. The boundary conditions of strings ending on such a D-brane preserve integrability \cite{Dekel:2011ja}, and we expect the 't~Hooft line  to define an integrable defect CFT.

The magnetic charges $q_i$ are quantized in half-integers due to the famous Dirac condition \cite{Dirac:1931kp} arising from consistency of  charged fields in the monopole background. One way to see why the monopole charge cannot be arbitrary (perhaps not the simplest) is to insist on spherical symmetry. That seems obvious for  (\ref{monopole}) but subtle quantum effects may break the symmetry unless extra conditions are imposed. Rotations are generated by the angular momentum operator, which for the monopole field was found in \cite{Fierz:1944}:
\begin{equation}
 L_i=-i\varepsilon _{ijk}x_jD_k-q\,\frac{x_i}{r}\,.
\end{equation}
This operator satisfies the $\mathfrak{su}(2)$ commutation relations $[L_i,L_j]=i\varepsilon _{ijk}L_k$ and commutes with the Hamiltonian\footnote{The Hamiltonian is the Klein-Gordon operator of a charged scalar in the monopole background. For vectors \cite{Olsen:1990jm,Weinberg:1993sg} and fermions \cite{Kazama:1976fm}, $\mathbf{L}$ does not commute with the Hamiltonian by itself. One needs to include spin, since only the total angular momentum is conserved.}:
\begin{equation}\label{Klein-G}
 K=-D_\mu ^2+\Phi _I^2=-\partial _t^2-D_i^2+\frac{q^2}{r^2}\,,
\end{equation}
where $D_i=\partial _i+iA_i$ is the covariant derivative in the field of the monopole. The scalar coupling comes from the $[\phi _I,\phi _J]^2$ term in the SYM Lagrangian, and combines nicely with the centrifugal energy  \cite{Fierz:1944} $[\ell(\ell+1)-q^2]/r$. The $q^2$ term cancels out giving rise to the conventional radial Hamiltonian of a free particle, as if the monopole were not present. The only influence of the background field on the radial motion is through the allowed range of $\ell$'s.

The mere existence of the angular momentum does not in itself guarantee rotational symmetry, $L_i$ must have a faithful representation in the Hilbert space. The $\mathfrak{su}(2)$ multiplets are then constructed  from the highest-weight state by raising and lowering with $L_\pm=L_1\pm iL_2$:
\begin{equation}
 L_3Y_{\ell\ell}=\ell Y_{\ell\ell},\qquad 
 L_+Y_{\ell\ell}=0,\qquad 
 Y_{\ell m}\propto L_-^{\ell-m}Y_{\ell \ell},
\end{equation}
thus defining the full set of spherical functions.
The tricky point in defining $L_i$ is the gauge choice. One (slightly unconventional) option is\footnote{Then $F=dA$ is the volume form on $S^2$ as required by (\ref{monopole}). In this gauge, the Dirac string pierces through the monopole in both directions. Normally one eliminates half of the string by a gauge transformation $A\rightarrow A\pm q\,d\varphi $, which makes the gauge potential regular on the upper/lower hemisphere. Our choice is more singular but also more symmetric, and has an advantage of leaving $L_3$ unaffected thus removing some unphysical phases in the Green's functions.} 
\begin{equation}
 A=-q\cos\theta \,d\varphi.
\end{equation}
The components of the angular momentum are then given by
\begin{equation}
 L_3=-i\partial _\varphi ,\qquad L_\pm=\,{\rm e}\,^{\pm i\varphi }
 \left(\pm\partial _\theta +i\cot\theta\,\partial _\varphi -\frac{q}{\sin\theta } \right).
\end{equation}
It is now straightforward to solve the highest-weight conditions. By simple integration we get:
\begin{equation}
 Y_{\ell\ell}=\,{\rm e}\,^{i\ell\varphi }\sin^l\theta \tan^q\frac{\theta }{2}\,.
\end{equation}
The key point is that this should be a regular  on the whole $S^2$.
The condition of regularity at $\theta =0$ and $\theta =\pi $ (such that $Y_{\ell\ell}\in L^\infty (S^2)$) requires that
\begin{equation}
 \ell\pm q\in\mathbbm{Z_+},
\end{equation}
which means that $\ell =q,q+1,\ldots $ and both $2q\in\mathbbm{Z_+}$ and $2l\in\mathbbm{Z}_+$. The lower harmonics are obtained by repeated application of $L_-$, and can be expressed through the Jacobi polynomials \cite{Tamm:1931dda}. It is important to notice that the monopole leads to fractionalization of the angular momentum: for the minimal magnetic charge $q=1/2$ the orbital angular momentum is quantized in {\it half-integers}: $\ell =1/2,3/2,\ldots $

The monopole harmonics $Y_{\ell m}(\theta ,\varphi )$ retain many properties of the usual spherical functions \cite{Wu:1976qk,Wu:1977qk}, and can be used to construct the eigenbasis of the Klein-Gordon operator (\ref{Klein-G}) and its cousins. The scalar propagator, for example, admits the following spectral representation:
\begin{equation}\label{sc-prop}
 D(x,x')=\frac{1}{8\pi ^2rr'}\left(\frac{1+\eta }{2}\right)^q\sum_{\ell=q}^{\infty }(2\ell+1)Q_\ell(\xi )P_{\ell-q}^{(0,2q)}(\eta  ),
\end{equation}
where $P_n^{(\alpha ,\beta )}(\eta )$ are the Jacobi polynomials normalized as  $P_n^{(0 ,\beta )}(1)=1$ and $Q_\ell(\xi )$ is the Legendre function of the second kind. The variables $\xi $ and $\eta $ are the standard conformal cross-ratios of two points and a line \cite{Buchbinder:2012vr}:
\begin{eqnarray}
 \xi&=&\frac{(t-t')^2+r^2+r'{}^2}{2rr'},
\nonumber \\
\eta &=&\frac{\mathbf{x}\cdot \mathbf{x}'}{rr'}\,.
\end{eqnarray}

Expansion in spherical harmonics can be alternatively viewed as a Kaluza-Klein reduction on the sphere, with the radial coordinate and time combining into $AdS_2$:
\begin{equation}
 ds^2=r^2\left(\frac{dt^2+dr^2}{r^2}+d\Omega _{S^2}^2\right).
\end{equation}
The cross-ratios $\xi $ and $\eta $ are simply related to the geodesic distances on $AdS_2$ and $S^2$, respectively. The function $Q_\ell(\xi )$ appearing in the spectral decomposition can be identified with the  $AdS_2$ propagator of a minimally coupled scalar of mass $m^2=\ell(\ell+1)$\footnote{The generic $AdS_d$ propagator is a hypergeometric function \cite{DHoker:2002nbb}. The latter is equivalent to the Legendre function when  specified to $d=2$.}. By the usual AdS/CFT logic each KK mode corresponds to a boundary operator of dimension $\Delta =\ell+1$. In the present context these are the defect operators, loosely equivalent to field insertions in the 't~Hooft line.

\subsection{Protected operators}

The propagators in the monopole background are necessary to develop perturbation theory for the 't~Hooft loop. Its complexity is comparable to perturbation theory  in AdS/CFT, the propagators are just the same. The simplest among them are the "anomalous" propagators that vanish in the absence of the monopole due to R-symmetry constraints. Consider, as an example, a complex scalar
\begin{equation}
 Z=(n_I+in'_I)\phi _I,
\end{equation}
where $n_I$ is the vector from the definition of the 't~Hooft loop and $n'_I$ is a unit six-vector orthogonal to it: $n\cdot n'=0$. 
 The longitudinal and transverse scalars enter with the opposite signs in the $\left\langle ZZ\right\rangle$ correlator, and tend to cancel one another. The cancellation is complete in vacuo, when the scalars have the same propagator. In the monopole background the propagators are different\footnote{The transverse scalar $n'\cdot \phi$ literally has the propagator (\ref{sc-prop}). The story is little bit more complicated for the longitudinal scalar $n\cdot \phi $ which mixes with the gauge fields \cite{Kristjansen:2023ysz}. The spin-$\ell$ term in its spectral decomposition contains three
terms with different $AdS$ masses:
 $Q_\ell$, $Q_{\ell-1}$ and $Q_{\ell+1}$  (see \cite{Kristjansen:2023ysz} for more details).}, but drastic simplifications still occur:
as shown in  \cite{Kristjansen:2023ysz} for $q=1/2$ and in the collinear limit $\eta =1$, the sum over modes telescopes leaving a simple finite residue\footnote{We use the notations of \cite{gradshteyn2014table} for special functions. These may sometimes differ from that of {\tt Mathematica}. The Legendre function $Q_\nu (z)$ is defined to be analytic for $\mathop{\mathrm{Re}} z>1$ in \cite{gradshteyn2014table} while {\tt Mathematica} treats $Q_\nu (z)$ as a functions with a cut from $1$ to $\infty $, as a result $Q_\nu (x\pm i0)^{{\rm Math}}=Q_\nu (x)^{\rm G.R.}\pm i\pi P_\nu (x)/2$. for real $x>1$. The argument of the elliptic integral in {\tt Mathematica} is $k^2$ while it is $k$ in \cite{gradshteyn2014table}.}:
\begin{equation}
 \left\langle Z_{1i}(x)Z_{j1}(x')\right\rangle=\delta _{ij}\,\frac{Q_{-1/2}(\xi )-Q_{1/2}(\xi )}{32\pi ^2rr'}=\delta _{ij}\,\frac{\mathop{\mathrm{Re}}\mathbf{E}}{16\pi ^2rr'}\,,
\end{equation}
where $\mathbf{E}\equiv \mathbf{E}(k)$ is the complete elliptic integral of the second kind with the modulus
\begin{equation}
 k^2=\frac{1+\xi }{2}\,.
\end{equation}
The color $i$ and $j$  indices transform under the unbroken $U(N-1)$. 
Importantly, the anomalous propagator has a finite OPE limit $\xi \rightarrow 1$:
\begin{equation}\label{ZZ}
 \left\langle Z_{1i}(x)Z_{j1}(x)\right\rangle=\delta _{ij}\,\frac{1}{16\pi ^2r^2}\,.
\end{equation}

We can use this result to calculate expectation values for chiral primaries:
\begin{equation}
 \mathcal{O}_{\rm CPO}=\mathop{\mathrm{tr}}Z^L.
\end{equation}
The leading order is equivalent to substituting the classical solution (\ref{monopole}) for $Z$ which gives $(2r)^{-L}$. The one-loop correction involes one $Z-Z$ propagator  evaluated at coincident points. From (\ref{ZZ}) we get:
\begin{equation}\label{NLO-CPO}
 \left\langle \mathcal{O}_{\rm CPO}\right\rangle=(2r)^{-L}\left(
 1+\frac{\lambda L}{4\pi ^2}+\ldots 
 \right).
\end{equation}
A comment on combinatorics is in order here. Only $Z_{1j}$ fields have a non-trivial two-point function and only $Z_{11}$ has a tree-level expectation value. Connecting a pair of $Z$s  by a propagator creates a loop with the color index  $1$ outside and $i\neq 1$ inside, but $\left\langle Z_{ij}\right\rangle=0$ and no fields should be left inside, meaning that only nearest neighbors can interact. That gives a combinatorial factor of $L$.

The loop corrections can be actually computed to all orders using localization of the path integral, worked out for the 't~Hooft loop in \cite{Gomis:2011pf}. We can also start (following \cite{Kristjansen:2023ysz}) with the known expectation value of a CPO in the presence of the Wilson line \cite{Semenoff:2001xp,Okuyama:2006jc} and then apply S-duality. The S-duality inverts the coupling: $\lambda \rightarrow 16\pi ^2N^2/\lambda $ and interchanges Wilson and 't~Hooft loops. That gives, in the planar approximation \cite{Kristjansen:2023ysz}\footnote{Our normalization conventions differ from those in  \cite{Kristjansen:2023ysz} where the operators are rescaled to have a unit two-point functions far away from the defect.}:
\begin{equation}
 \left\langle \mathcal{O}_{\rm CPO}\right\rangle=(4r)^{-L}\left[
 \left(\sqrt{1+\frac{\lambda }{\pi ^2}}+1\right)^L-
  \left(\sqrt{1+\frac{\lambda }{\pi ^2}}-1\right)^L
 \right].
\end{equation}
Expanding in $\lambda $ we recover (\ref{NLO-CPO}).

\subsection{$SO(6)$ sector}

As for non-protected operators, their one-point functions are way more complicated but amenable to integrability methods quite similar to those for the domain wall defects. We shall discuss a few particular cases that illustrate the general approach (yet to be developed), restricting to the tree-level one-point functions obtained by replacing constituent fields of the operator by the classical background (\ref{monopole}).
The result can be always represented as an overlap of the spin-chain wavefunction, e.g. $\Psi $ from (\ref{so6-operators}), with a fixed boundary state. If we are lucky and this boundary state has special integrability properties, its overlaps are given by a determinant expression similar to the ones encountered in secs.~\ref{SYM} and \ref{ABJM}.

For the scalar operators in the $SO(6)$ sector, of generic form (\ref{so6-operators}), the tree-level expectation value is the overlap with a tensor product state:
\begin{equation}
 C^{SO(6)}_{\rm tHooft}(\mathbf{u})=2^{-L}\frac{\left\langle {\rm TPS}\right.\!\left| \mathbf{u}\right\rangle}{\left\langle \mathbf{u}\right.\!\left| \mathbf{u}\right\rangle^{\frac{1}{2}}}\,,\qquad 
 {\rm TPS}^{I_1\ldots I_L}=n^{I_1}\ldots n^{I_L}.
\end{equation}
In other words, the one-point function measures a particular component of the wavefunction, for instance $\Psi _{1\ldots 1}$. This state is integrable and its overlaps admit the following determinant representation  \cite{deLeeuw:2019ebw}:
\begin{equation}
 \frac{\left\langle {\rm TPS}\right.\!\left| \mathbf{u}\right\rangle}{\left\langle \mathbf{u}\right.\!\left| \mathbf{u}\right\rangle^{\frac{1}{2}}}
 =2^{-\frac{L}{2}}\sqrt{\frac{Q_2(0)Q_2\left(\frac{i}{2}\right)}{Q_1(0)Q_1\left(\frac{i}{2}\right)Q_3(0)Q_3\left(\frac{i}{2}\right)}\,\mathop{\mathrm{Sdet}}G}
 \,.
\end{equation}
The existence of a determinant formula signals integrability of the boundary state thus confirming expectations that the 't~Hooft line defines an integrable dCFT.

\subsection{$SL(2)$ sector}

Operators with spin can  acquire expectation values only for defects of co-dimension greater than one. To illustrate this new feature we consider operators obtained from the chiral primary $\mathop{\mathrm{tr}}Z^L$ by dressing it with the light-cone derivatives, $D\equiv D_0+D_3$ (in Minkowski signature):
\begin{equation}\label{slops}
 \mathcal{O}=\sum_{\left\{n_\ell\right\}}^{}\frac{1}{n_1!\ldots n_L!}\,
 \Psi _{n_1\ldots n_L}
 \mathop{\mathrm{tr}}D^{n_1}Z\ldots D^{n_L}Z.
\end{equation}
An operator with $S=n_1+\ldots +n_L$ derivatives has spin $S$, twist $L$ and bare dimension $L+S$.

It is often convenient to consider the generating function instead of the local operators. For a single field the generating function is
\begin{equation}\label{genfunc}
 UZ(x+\xi v)U^{-1}=\sum_{n=0}^{\infty }\frac{\xi ^n}{n!}\,D^nZ(x),
\end{equation}
where $v^\mu =(1,0,0,1)$ and $U$ is the light-like Wilson line:
\begin{equation}
 U={\rm P}\exp\left(\int_{0}^{\xi }d\eta \,v^\mu A_\mu (x+\eta v)\right).
\end{equation}
The generating function for generic operators is a light-ray Wilson line with $L$ insertions of $Z$. The dilatation  operator in the $\mathfrak{sl}(2)$ sector can be reformulated entirely in terms of the light-ray Wilson lines \cite{Belitsky:2003ys}. 

The expectation values in the monopole background at the leading order are obtained by substituting the classical background (\ref{monopole}) into the operator at hand. The classical field is Abelian and time-independent, so the Wilson line cancels and the light-cone shift effectively reduces to spacial translations along $\mathbf{v}=(0,0,1)$. Comparing the Legendre expansion \cite{Legendre:1785}:
\begin{equation}
 \frac{1}{|\mathbf{x+\xi \mathbf{v}}|}=\sum_{n=0}^{\infty }\frac{(-\xi )^n}{r^{n+1}}\,P_n(\cos\theta ),
\end{equation}
with (\ref{genfunc}) we can read off the boundary state:
\begin{equation}
 \left|B(\theta )\right\rangle=\sum_{n=0}^{\infty }(-1)^nP_n(\cos\theta )\left|n\right\rangle.
\end{equation}
This gives an overlap representation for the one-point function:
\begin{equation}
 \left\langle \mathcal{O}\right\rangle=\frac{1}{2^Lr^{L+S}}\,\,\frac{\left\langle {\rm Bsl}\right.\!\left|\mathbf{u} \right\rangle}{\left\langle \mathbf{u}\right.\!\left|\mathbf{u} \right\rangle^{\frac{1}{2}}}\,,
\end{equation}
where $\left|{\rm Bsl}\right\rangle$ is again of the tensor-product form:
\begin{equation}
 \left|{\rm Bsl}\right\rangle=\left|B\right\rangle\otimes\ldots \otimes\left|B\right\rangle.
\end{equation}

Overlaps with the eigenstates of the dilatation operator admit a determinant representation:
\begin{equation}
 \frac{\left\langle {\rm Bsl}\right.\!\left|\mathbf{u} \right\rangle}{\left\langle \mathbf{u}\right.\!\left|\mathbf{u} \right\rangle^{\frac{1}{2}}}
 = (\sin\theta )^S\sqrt{\frac{Q(0)}{Q\left(\frac{i}{2}\right)}\,\mathop{\mathrm{Sdet}}G}\,.
\end{equation}
This formula was conjectured \cite{Kristjansen:2023ysz} on account of structural similarity to other integrable boundary states and was rigorously proven in \cite{Gombor:2023bez}. Noticeably, its prefactor vanishes at $\theta =0$, unless $S=0$. This follows from the representation theory of $\mathfrak{sl}(2)$. The Bethe eigenstates $\left|\mathbf{u}\right\rangle$ are $\mathfrak{sl}(2)$ primaries annihilated by the superconformal generator: $K\left|\mathbf{u}\right\rangle=0$. Since $P_n(1)=1$, 
\begin{equation}
 \left|B(0)\right\rangle_\ell=\sum_{n=0}^{\infty }(-1)^n\left|n\right\rangle_\ell
 =\sum_{n=0}^{\infty }\frac{(-1)^n}{n!}P_\ell^n\left|n\right\rangle_\ell
 =\,{\rm e}\,^{-P_\ell}\left|0\right\rangle_\ell,
\end{equation}
where we explicitly label the position $\ell$ in the tensor product. The total tensor product is thus a vacuum descendant:
\begin{equation}
 \left|{\rm Bsl}(0)\right\rangle=\,{\rm e}\,^{-P}\left|0\right\rangle.
\end{equation}
Since $P^\dagger =K$, we find that $\left\langle {\rm Bsl}(0)\right.\!\left| \mathbf{u}\right\rangle=0$ unless $\mathbf{u}$ is the trivial ground state.

\subsection{Gluon operators}

Another set of operators with spin, forming a closed sector under renormalization\footnote{At one loop.}, is constructed from the (anti-)self-dual components of the field strength, which we single out with the help of the 't~Hooft symbols \cite{tHooft:1976snw}:
\begin{equation}
 F^\pm_i=\frac{1}{2}\eta ^\pm_{i\mu \nu }F^{\mu \nu },
 \qquad 
 \eta _{i\mu \nu }=\varepsilon _{0i\mu \nu }\pm\delta _{0\mu }\delta _{i\nu }
 \mp\delta _{0\nu }\delta _{i\mu }.
\end{equation}
The linear combinations of the form
\begin{equation}
 \mathcal{O}=\Psi ^{i_1\ldots i_L}\mathop{\mathrm{tr}}F_{i_1}^\pm\ldots F_{i_L}^\pm
\end{equation}
can be interpreted as states in the quantum $SU(2)$ spin chain of spin one\footnote{The 't~Hooft symbols decompose the field strength into irreducible $(\mathbf{3},0)\oplus(0,\mathbf{3})$  representations of $SO(4)\simeq SU_L(2)\times SU_R(2)$.}.

The boundary state is a tensor product:
\begin{equation}
 {\rm Bgl}_{i_1\ldots i_L}=x_{i_1}\ldots x_{i_L},
\end{equation}
whose overlaps with the spin-chain eigenstates again admit a determinant representation \cite{Kristjansen:2023ysz}:
\begin{equation}
 \frac{\left\langle {\rm Bgl}\right.\!\left|\mathbf{u} \right\rangle}{\left\langle \mathbf{u}\right.\!\left|\mathbf{u} \right\rangle^{\frac{1}{2}}}
 =\left(\frac{r}{\sqrt{2}}\right)^L(\sin\theta )^{S}
 \sqrt{\frac{Q(0)}{Q\left(\frac{i}{2}\right)}\,\mathop{\mathrm{Sdet}}G}\,,
\end{equation}
where $S$ is the $SU(2)$ spin.

\section{Summary and Outlook\label{conclusion}}

We have reviewed three holographic set-ups where the introduction of a defect breaks half of the supersymmetries  and
limits conformal symmetry to a sub-space but where nevertheless integrability is preserved. Whereas all-loop asymptotic
integrability has only been explicitly demonstrated for the Nahm pole version of ${\cal N}=4$ SYM theory, it is expected to 
hold for the domain wall version of ABJM theory and for the 't Hooft loop in ${\cal N}=4$ SYM theory as well.  Other Nahm
pole versions of ${\cal N}=4$ SYM with no supersymmetry are known to be integrable only at the leading order~\cite{DeLeeuw:2019ohp,Gombor:2020kgu} or not integrable at all~\cite{deLeeuw:2019sew}.  For ABJM theory, no other Nahm pole solutions apart from the one presented here are known but it is an interesting open question whether monopole operators  could be described in terms of integrable boundary states in the same way as the 't Hooft line in ${\cal N}=4$ SYM theory. It would likewise be interesting to investigate whether there exists integrable defects of even co-dimension in any of these theories. All of the known integrable cases  correspond to defects of odd co-dimension. 

Finally,  progress on integrability of lower dimensional versions of the AdS/CFT correspondence~\cite{Babichenko:2009dk,Sfondrini:2014via,Sorokin:2011rr,Hoare:2014kma}
opens an avenue for revealing novel types of  holographic integrable defect CFTs.

\section*{Acknowledgements.}
The authors wish to thank Manu Paranjape for the invitation to the Conference in honor of Gordon Semenoff.
C.K.\ was supported by DFF-FNU through grant number 1026-00103B. K.Z. was supported by VR grant 2021-04578. 

\appendix
\section{On  spin chain integrability}
Good conformal operators of ${\cal N}=4$ SYM and of ABJM theory correspond to eigenstates of integrable spin chains with underlying symmetry algebras respectively $\mathfrak{psu}(2,2|4)$ and
$\mathfrak{osp}(6|4)$~\cite{Minahan:2002ve,Minahan:2008hf}. 
The starting point of AdS/CFT integrability is the fact that spin chains with nearest neighbor interactions and with 
these symmetries, or more generally symmetries derivable from $\mathfrak{gl}(N|M)$ symmetry  are  solvable by means of the Bethe ansatz approach~\cite{Saleur:1999cx}.
The choice of Dynkin diagram of the symmetry algebra (which is not unique for super Lie algebras) plays an important role in this description. An eigenstate of the spin chain in 
question is characterized by a set of Bethe roots $u_{a,j}$ where $a$ is an index that numbers the nodes of the Dynkin diagram and $j$ is an index which simply counts the number of
roots associated to a particular node, i.e. $j=1,\ldots,K_a$. Knowing the Bethe roots one can explicitly write down the wave function of the eigenstate and likewise the expression for its
corresponding conformal operator. The Bethe roots have to fulfil the following equations to produce an eigenstate:
\begin{eqnarray}\label{BAEs}
 \left(\frac{u_{a,j}-\frac{iq_a}{2}}{u_{a,j}+\frac{iq_a}{2}}\right)^L
 \prod_{b,k}\frac{u_{a,j}-u_{b,k}+\frac{iM_{ab}}{2}}{u_{a,j}-u_{b,k}-\frac{iM_{ab}}{2}}\equiv \,{\rm e}\,^{i\chi _{a,j}}=-1,\nonumber 
\end{eqnarray}
where $M_{ab}$ is the Cartan matrix of the symmetry algebra, and $a, b$ hence run from 1 to $n$ with $n$ being the number of nodes of the Dynkin diagram. Furthermore, $q_a$ 
consists of the labels of the representation in which the spin variable of the chain lives.  A root $u_{a,j}$ belonging to a node $a$ for which $q_a\neq 0$ is said to be momentum carrying.
A convenient way of representing a Bethe eigenstate is via the associated Baxter polynomials
\begin{equation}
Q_a(u)=\prod_{j=1}^{{K_a}}\left(u-u_{a,j}\right),
\end{equation}
with $a=1,\ldots,n$, $j=1,\ldots, K_a$.
The norm of a Bethe eigenstate is encoded in the determinant of the so-called Gaudin matrix~\cite{Gaudin:1983}  which is defined as follows 
\begin{equation}
 G_{aj,bk}=\frac{\partial \chi _{a,j}}{\partial u_{b,k}}\,.
\end{equation}
A boundary state is just a particular spin chain state, which is not a simple linear combination
of eigenstates, such as e.g.\ the matrix product state defined by the wave function in equation~(\ref{MPSk}). An interesting question is when a boundary state can be characterized as being integrable meaning in this connection that its overlaps with the Bethe eigenstates can be found in closed form. Generalizing the concept of an integrable boundary state from 
field theory~\cite{Ghoshal:1993tm} one can argue that an integrable spin chain boundary state must be a state which constitutes a  superposition of pairs of excitations with momenta $+p$ and $-p$~\cite{Piroli:2017sei}. This translates into the statement that the Bethe roots at the momentum carrying nodes must come in pairs with  opposite sign. This can happen in two different ways which
are denoted as chiral or a-chiral~\cite{Gombor:2020kgu}. In the chiral case the roots at a single momentum carrying node are paired with roots from the same node. In the a-chiral case roots from one
node are paired with roots from another node which appears in a symmetric way in the Dynkin diagram.  The SO$(6)$ overlaps calculated on the basis of the Dynkin 
diagram~in Fig.\ \ref{SO6Dynkin}
are chiral whereas the SU$(4)$ overlaps calculated on the basis of the Dynkin diagram in Fig.\ \ref{SU4Dynkin} are a-chiral. In both the chiral and the a-chiral cases the collection of roots has a $Z_2$ symmetry which can be used to define the super determinant of the Gaudin matrix, SdetG,~\cite{Kristjansen:2020mhn}, which appears in the various overlap formulas. 
Special care is needed to define the Gaudin super determinant  if there are Bethe roots at zero or at 
$\pm i/2$. Likewise, 
in overlap formulas singular roots should be excluded from the Baxter polynomials~\cite{Brockmann:2014b,deLeeuw:2016umh,Kristjansen:2021xno}.

\bibliographystyle{nb}

\end{document}